\documentclass[11pt]{article}

\usepackage{fullpage}

\usepackage[utf8]{inputenc}
\usepackage{siunitx}
\usepackage{xspace}
\usepackage{balance}
\usepackage{tikz}
\usepackage{pgfplots}
\usepackage{booktabs}
\usepackage{authblk}
\usepackage[ruled,vlined]{algorithm2e}
\usetikzlibrary{calc}

\newcommand{\children}{\texttt{children}\xspace} %ora ho incluso xspace
\newcommand{\maxw}{\texttt{max-w}\xspace}

\usepackage{amsmath,amssymb,amsfonts}
\usepackage{algorithmic}
\usepackage{graphicx}
\usepackage{textcomp}
\usepackage{xcolor}
\usepackage{hyperref}
\def\BibTeX{{\rm B\kern-.05em{\sc i\kern-.025em b}\kern-.08em
    T\kern-.1667em\lower.7ex\hbox{E}\kern-.125emX}}
\pgfplotsset{compat=1.14}
    
\begin{document}

\title{Zuckerli: A New Compressed Representation for Graphs}

\iffalse
\author{\IEEEauthorblockN{Anonymous}
\IEEEauthorblockA{\textit{Anonymous} \\
\textit{Anonymous}\\
City, Country \\
email address or ORCID}
\iffalse
\and
\IEEEauthorblockN{Given Name Surname}
\IEEEauthorblockA{\textit{dept. name of organization (of Aff.)} \\
\textit{name of organization (of Aff.)}\\
City, Country \\
email address or ORCID}
\and
\IEEEauthorblockN{Given Name Surname}
\IEEEauthorblockA{\textit{dept. name of organization (of Aff.)} \\
\textit{name of organization (of Aff.)}\\
City, Country \\
email address or ORCID}
\fi
}
\fi

\author[1,2]{Luca Versari\thanks{veluca@google.com}}
\author[2]{Iulia M. Comsa\thanks{iuliacomsa@google.com}}
\author[1]{Alessio Conte\thanks{conte@di.unipi.it}}
\author[1]{Roberto Grossi\thanks{grossi@di.unipi.it}}
\affil[1]{University of Pisa}
\affil[2]{Google Research, Z\"urich}

\date{}

\pgfdeclarelayer{bg}
\pgfsetlayers{bg,main}

\maketitle

\begin{abstract}
    Zuckerli is a scalable compression system meant for large real-world graphs. 
    %The last decade has witnessed a dramatic growth in size and numbers of the web and computer systems in general, leading to the availability of ever more and larger collections of data.  
    Graphs are notoriously challenging structures to store efficiently due to their linked nature, which makes it hard to separate them into smaller, compact components. Therefore, effective compression is crucial when dealing with large graphs, which can have billions of nodes and edges. Furthermore, a good compression system should give the user fast and reasonably flexible access to parts of the compressed data without requiring full decompression, which may be unfeasible on their system.
    
    Zuckerli improves multiple aspects of WebGraph, the current state-of-the-art in compressing real-world graphs, by using advanced compression techniques and novel heuristic graph algorithms. It can produce both a compressed representation for storage and one which allows fast direct access to the adjacency lists of the compressed graph without decompressing the entire graph. We validate the effectiveness of Zuckerli on real-world graphs with up to a billion nodes and 90 billion edges, conducting an extensive experimental evaluation of both compression density and decompression performance. We show that Zuckerli-compressed graphs are 10\% to 29\% smaller, and more than 20\% in most cases, with a resource usage for decompression comparable to that of WebGraph.

\end{abstract}

\iffalse
\begin{IEEEkeywords}
graphs, compression, compressed data structures
\end{IEEEkeywords}
\fi

\section{Introduction}

Graph compression essentially boils down to compressing the adjacency lists of a graph $G=(V,E)$, where its nodes are suitably numbered from 1 to $n=|V|$, and the adjacency list storing the neighbors of each node is seen as the sorted sequence of the corresponding integers from $[1, 2, \ldots, n]$. It is straightforward to use a $64$-bit word of memory for each integer (i.e.~edge), plus $O(n)$ words for the degrees and the pointers to the $n$ adjacency lists, thus requiring $O(n+m)$ words of memory for the standard representation of $G$. 

The challenge is to use very few bits per edge and node, so as to squeeze $G$ into as little space as possible. This can make a dramatic difference for massive graphs, particularly if the compressed graph fits into main memory, while its standard representation does not. The over $450$ bibliographic entries in a recent survey on lossless graph compression~\cite{besta2018survey} give a measure of the increasing interest for this line of research. Among the numerous proposals, the WebGraph framework~\cite{boldi2004webgraph,boldi2004webgraph2} is widely recognized as the touchstone for its outstanding compression ratio.

In this paper, we consider the lossless graph compression scenario, showing how to compress $G$ and supporting two kinds of operations on the resulting compressed representation of $G$:

\textbf{Full decompression}: Decompress the representation entirely, obtaining the standard representation of $G$.

\textbf{List decompression}: For any given node $u \in [n]$, decompress incrementally the adjacency list of $u$, while keeping the rest compressed.

List decompression can allow us to run some graph algorithms directly on the compressed representation on the graph: several fundamental algorithms, such as a graph traversal, are based on partially scanning adjacency lists that are
%, and thus just what is needed inside each list is 
decompressed during the scan. 

% Note that list decompression is needed to be incremental in some algorithms, such as a graph traversal, where some adjacency lists are partially scanned and thus just what is needed inside each list is decompressed during its scan. (This notion is called lazy iterators in WebGraph). 

On the other hand, we do not want to support decompressing a single edge (i.e. directly checking adjacency between two nodes) for two reasons: it degrades the performance of scanning an adjacency list, and many of the well-known graph algorithms hardly require to access few random items of an adjacency list without accessing the list from the beginning. 
%Moreover, as most nodes in large real-world graphs typically have very small adjacency lists (i.e., in double digits), any attempt to jump parts of it would just degrade the performance due to the extra machinery required. 
% Moreover, as we will see in our experiments, 
Moreover, scanning a list is so fast in our implementation that any attempt to jump parts of it would just degrade the performance due to the extra machinery required. 

In this paper, we present a new graph compressor called Zuckerli. By incorporating advanced compression techniques and novel heuristic algorithms, Zuckerli is able to replace Webgraph-compressed graphs with a compressed structure representing the same data, but that uses 20\% to 30\% less space for web graphs, and 10\% to 15\% less space for social networks, saving significant space on storage media. These savings also hold when compressing a graph for list decompression, compared to the corresponding list decompression mode of WebGraph. Decompression is highly tuned and very fast, providing millions of edges per second on a commodity computer. 

To the best of our knowledge, Zuckerli significantly improves the state-of-the-art in graph compression when full or list decompression is supported.

\medskip
\noindent\textbf{Related work.}
Compressing graphs is a well-studied problem. The WebGraph framework~\cite{boldi2004webgraph,boldi2004webgraph2} exploits two well known properties shared by web graphs (and, in a smaller measure, by social networks), \emph{locality} and \emph{similarity}, originally exploited by the LINKS database~\cite{randall2002link}. WebGraph is the graph compression technique most directly related to Zuckerli, as it uses the above properties.

More recently, an approach called Log(Graph) and based on \emph{graph logarithmization}~\cite{besta2018log} has been explored. The analysis conducted shows that, while Log(Graph) achieves better performance while performing various operations, the WebGraph framework is still the most competitive approach in terms of compression ratio, especially for web graphs.

Another well-known approach to graph compression are $k^2$-trees~\cite{brisaboa2009k}, which use a succinct representation of a bidimensional $k$-tree on the adjacency matrix of the graph. Unlike WebGraph, this scheme allows for accessing single edges, without requiring the decoding of full adjacency lists at a time. As a consequence, it achieves somewhat worse compression ratios, but is more suited for applications where single edges are queried. The $k^2$-trees have been subsequently improved by $2$D block trees~\cite{brisaboa2018two}, a LZ77-like approach that can compress bidimensional data. As with $k^2$-trees, it allows for querying single edges; however, it achieves significantly improved compression ratios, at the cost of a hit in query time. A brief experimental comparison between Zuckerli, $k^2$-trees and $2$D block trees can be found in Section~\ref{sec:experiments}.

Some other approaches follow a different philosophy, that is, providing access to the compressed graph with a wide range of complex operations, or even a query language, at the cost of sub-optimal compression ratios. This is the case for example of ZipG~\cite{zipg2017}, a distributed graph storage system aims at compactly storing a graph, including semantic information on its nodes and edges, while allowing access to this information via a minimal but rich API. We refer the reader to the survey in~\cite{besta2018survey} for a panoramic view of the research on graph compression.

\medskip

The paper is organized as follows. Section~\ref{sec:entropy} discusses some methods to encode integers, which are at the heart of our compression algorithms and are used to encode all the data that results from the higher-level compression scheme. Section~\ref{sec:zuckerli} describes the Zuckerli high-level encoding scheme, which, in brief, consists in \emph{block-copying}, that is re-using parts of the adjacency lists of previous nodes to encode the adjacency list of current nodes, \emph{delta-coding} of values that are not copied and \emph{context-modeling} of all the values to improve compression. This section also describes \emph{heuristics} to improve the encoding choices made by the encoder. We then report the experimental study in Section~\ref{sec:experiments}, and draw conclusions in Section~\ref{sec:conclusions}.

\section{Encoding Integers}
\label{sec:entropy}

Our graph compression method modifies the adjacency lists, which are sequences of integers, to produce other sequences of integers that can be encoded more succinctly. Thus, encoding methods for the integers are at the heart of Zuckerli, and we discuss the ones that we employ from existing literature, or that we design for this purpose.

\subsection{Multi-context entropy encoding}

Zuckerli uses Huffman coding~\cite{huffman1952method} when list decompression is supported, and Asymmetric Numeral Systems (ANS)~\cite{duda2009asymmetric} when full decompression is required only. 

Conceptually, ANS encodes a sequence of input symbols in a single number that can be represented with a number of bits that is close to the entropy of the data stream. Thus, it is a form of arithmetic coding (whose idea goes back to Shannon~\cite{shannon-48}), but compared to traditional methods of arithmetic coding it can achieve faster compression and decompression speeds. The encoding process adds a symbol $s$ to the stream represented by $x$ by producing a new integer $C(s,x) = M \lfloor x/F_s \rfloor + B_s + (x \bmod F_s)$, where $M$ is the sum of the frequencies of all the symbols, $F_s$ is the frequency of the symbol $s$ and $B_s$ is the cumulative frequency of all the symbols before $s$. This function is invertible, hence the decoder can reverse this process and produce the stream of symbols starting from $x$.

Like all variants of arithmetic coding, practical implementations of ANS do not use arbitrary precision arithmetic, but rather they keep an internal state in a fixed range $[S, 2^b S)$ that is manipulated for each symbol in the stream: when the state overflows, it yields $b$ bits during encoding; when the state underflows, it consumes $b$ bits when decoding. For correct decoding, it is required that $S$ is a multiple of $M$. In our case, we set $S = 2^{16}$, $M = 2^{12}$, and $b = 16$. Since the decoding procedure is just the reverse of the decoding procedure, ANS makes it easy to interleave non-compressed bits.

The variant of ANS used by Zuckerli is inspired by the one employed in the new standard JPEG~XL~\cite{10.1117/12.2529237} for lossy image compression.

When list decompression is supported, one disadvantage of ANS (as well of as other encoding schemes that can use a non-integer number of bits per encoded symbol) is that it requires keeping track of its internal state. For decoding to successfully be able to resume from a given position in the stream, it is also necessary to be able to recover the state of the entropy coder at that point of the stream, which would cause significant per-node overhead if using ANS. Thus, in this case, Zuckerli switches to using Huffman coding.

Huffman coding represents every input symbol with a variable number of bits, without having an internal state. The bits of the representation are chosen in such a way that no two symbols share the same prefix of bits (to allow to decode correctly). As a consequence, Huffman coding easily allows seeking, but cannot use less than one bit per symbol.

Both Huffman and ANS use a \emph{context} or model, which is a prediction of the probability distribution for the symbols in the stream that are obtained from the adjacency lists. The more accurate the prediction is, the closer to optimal the compression gain will be. As both the encoder and the decoder must share the same context, Zuckerli has to store the probability distributions corresponding to a context when encoding the graph. Symbols to be encoded are spread among multiple contexts, allowing more
precise encoding when symbols are assumed to belong to different probability distributions. Hence, \emph{multi-context entropy coding} is one significant source of improvements of Zuckerli in comparison to other approaches.

\subsection{Hybrid integer encoding}
\label{sub:hybrid}
When compressing streams, both Huffman and ANS encode the symbols belonging to a given alphabet and thus benefit from having a reduced alphabet size. However, alphabet may grow too large in our case as Zuckerli needs to encode integers of arbitrary length and cannot use a distinct symbol for each integer. Zuckerli thus introduces a new \emph{hybrid integer encoding} scheme, described below. This generalizes a scheme that was initially developed for image compression in JPEG XL~\cite{10.1117/12.2529237}\footnote{In particular, the scheme described in~\cite{10.1117/12.2529237} corresponds to the scheme described here with $k=4, i=1, j=0$.}.

Zuckerli's hybrid encoding scheme is defined by three parameters: $i$, $j$ and $k$, with $k \ge i + j$ and $i, j \ge 0$.  Every integer in the range $[0, 2^k)$ is encoded directly as symbol in the alphabet. 

Any other integer $x \geq 2^k$ is encoded as follows. First, consider the binary representation of $x$: $b_p  b_{p-1} \cdots b_1$, where $b_p=1$ is the highest non-zero bit. Equivalently, identify $x$ with its corresponding triple $(m, t, l)$ where $m$ is the integer formed by the $i$ bits $b_{p-1}\cdots b_{p-i}$ following $b_p$, $l$ is the integer formed by the rightmost $j$ bits $b_j\cdots b_1$, and $t$ is the integer encoded by the bits between those of $m$ and $l$, as illustrated below:
\[1\overbrace{b_{p-1}\dots b_{p-i}}^{m}\overbrace{b_{p-i-1}\dots b_{j+1}}^{t}\overbrace{b_j\dots b_1}^{l} \]
Clearly, given the triple $(m, t, l)$, we can reconstruct $x$. We conveniently encode that triple by a pair $(s,t)$ where $s = 2^k + (p-k-1) \cdot 2^{i+j} + m \cdot 2^j + l$ encodes, respectively, the value of $k$ by $2^k$, the value of $p \geq k+1$ by $(p-k-1) \cdot 2^{i+j}$, the value of $m$ as $m \cdot 2^j$ followed by $l$.

For example, for $k=4$, $i=1$, and $j=2$, the integer $x=105$ has binary representation $\mathtt{1}\,\mathtt{1}\,\mathtt{0100}\,\mathtt{11}$ and its corresponding triple is $(1, 4, 3)$,  and thus encoded as the pair $( 16+ 3 \cdot 8 + 1 \cdot 4 + 3 , 4) = (47,4)$ where $p=8$. As another example, when $k=4$, $i=1$ and $j=1$, the integers from $0$ to $15$ are encoded with their corresponding symbol $s$ in the alphabet, and $t$ is empty; $23$ has binary representation $\mathtt{10111}$ and thus is encoded as symbol $\mathtt{17}$ (the highest set bit is in position $5$, the following bit is $0$, and the last bit is $1$), followed by the two remaining bits $\mathtt{11}$; $33$ is encoded as symbol $\mathtt{21}$ (highest set bit is in position $6$, following bit is $0$ and last bit is $1$) followed by the three remaining bits $\mathtt{000}$.

\iffalse
$0 \mapsto (\mathtt{0}, \ )$, 
$1 \mapsto (\mathtt{1}, \ )$,~\dots, 
$15 \mapsto (\mathtt{15}, \ )$. 

Integers larger than $15$ are encoded as follows:
$16 \mapsto (\mathtt{16}, \mathtt{00})$,
$17 \mapsto (\mathtt{17}, \mathtt{00})$,~\dots, 
$23 \mapsto (\mathtt{17}, \mathtt{11})$,
$24 \mapsto (\mathtt{18}, \mathtt{00})$,~\dots,
$31 \mapsto (\mathtt{19}, \mathtt{11})$, 
$32 \mapsto (\mathtt{20}, \mathtt{000})$, 
$33 \mapsto (\mathtt{21}, \mathtt{000})$,~\dots
\fi

The advantage of this scheme is that $s$ has a smaller range than $x$, and can thus be entropy-encoded by either Huffman or ANS: using this representation, $r$-bit integers require at most $2^k + (r-k-1) \cdot 2^{i+j}$ symbols in the alphabet instead of $2^r$. %Table~\ref{tab:hyb} reports the effects of choosing $k$ and $i$ in our experiments (see Section~\ref{sec:experiments} for more details).

As for $t$, it is stored as-is in the encoded file, just after entropy coding $s$. Note that it is possible to compute the number of bits of $t$ from $s$, without knowing $x$: this allows the decoder to know how many bits to read. The procedure to decode an integer from the $(s, t)$ pair consists of recovering the corresponding triple $(m, t, l)$ and then reconstructing $x$. The procedure is detailed in Algorithm~\ref{alg:hyb}.

\begin{algorithm}[t]
\SetAlgoLined
  \If{$s < 2^k$} {
    \Return $s$\;
  }
  $l \gets (s - 2^k) \mod 2^j$\;
  $s \gets \frac {s - l - 2^k}{2^j}$\;
  $m \gets s \mod 2^i$\;
  $n \gets \frac {s - m}{2^i}$   \qquad (note $n = p-k-1$)\;
  \Return $2^{n+k} + m \cdot 2^{n+k-i} + t \cdot 2^{j} + l$\;
  \caption{How to decode an $(s, t)$ pair.\label{alg:hyb}}
\end{algorithm}

\subsection{Negative integers}

We encode a negative integer $s$ as follows, as it is easy to reverse this bijection between integers and natural numbers~\cite{boldi2004webgraph}.
\begin{equation}
  \label{eqn:packsigned}
  x \rightarrow
  \begin{cases}
    2\cdot x & \text{if } x \ge 0 \\
    -2\cdot x - 1 & \text{if } x < 0 
  \end{cases}
\end{equation}

\section{Graph compression in Zuckerli}
\label{sec:zuckerli}

This section details the graph compression scheme used by Zuckerli.
%This
%scheme is based on two successful ideas from WebGraph: 
%\begin{enumerate*}[label=(\roman*)]
%    \item nodes are numbered according to the layered label propagation algorithm~\cite{BoldiRSV11}, and
%    \item locality and similarity are exploited in the adjacency lists of real-world networks.
%\end{enumerate*}
%~\cite{boldi2004webgraph}, 
%a state-of-the-art solution for web graph compression, 
%We first present a brief summary. 

\subsection{Brief summary of WebGraph}
\label{sec:webgraph}

As Zuckerli reuses and improves on multiple aspects of WebGraph, here we provide a brief summary of the WebGraph scheme.

Let $W$ and $L$ be global parameters representing the ``window size'', which is limited to speed up compression time, and the ``minimum
interval length''. For each node $u \in V$, WebGraph encodes its \emph{degree} $\deg(u)$ and, if $\deg(u) >0$, the following information for the adjacency list of $u$:
\begin{enumerate}
  \item \label{item:reference} A \emph{reference
    number} $r$, which can be either a number in $[1, W)$, meaning that the
    list is represented by referencing the adjacency list of node $u-r$ (called
    \emph{reference list}), or $0$, meaning that the list is represented without
    referencing any other list.
  \item \label{item:blocks} If $r>0$, it is followed by a list of integers indicating
    the indices where the reference list should be split to obtain contiguous
    \emph{blocks}. Blocks in even positions represent edges that should be
    copied to the current list. The format contains, in this order, the number
    of blocks, the length of the first block, and the length minus $1$ of all
    the following blocks (since no block except the first may be empty). The
    last block is never stored, as its length can be deduced from the length of
    the reference list.
  \item \label{item:intervals} A list of \emph{intervals} follows; each interval 
    has at least $L$ consecutive nodes that are not copied from the blocks in point~\ref{item:blocks}.
  \item \label{item:residuals} Whatever nodes are left from points~\ref{item:blocks}--\ref{item:intervals} are called \emph{residuals}, and they are \emph{delta-coded}. Their number can be deduced by the degree, the number of copied edges and the number of edges represented by intervals. The first residual is encoded by difference with respect to $u$ (and thus it can be a negative number), and each of the remaining residuals is represented by difference with respect to the previous residual, minus $1$.
\end{enumerate}

WebGraph represents the resulting sequence of non-negative integers by
using $\zeta$ codes~\cite{boldi2004webgraph2}, a set of universal codes
particularly suited to represent integers following a power-law distribution.

Moreover, to guarantee fast access to individual adjacency lists, WebGraph limits the length
of the \emph{reference chain} of each node. In particular, a reference chain is
a sequence of nodes $u_1, \dots, u_\ell$ such that node $n_{i+1}$ uses node $n_i$
as a reference $r$. 
%$\ell$ is then the length of the reference chain. 
%WebGraph guarantees that e
Every chain has length $\ell \leq R$, where $R$ is a
global parameter.

\subsection{Zuckerli scheme}
\label{sec:scheme}

In this section, we summarize the novel aspects introduced by Zuckerli in relation to WebGraph.

First, Zuckerli entropy-encodes the integers, as described in Section~\ref{sec:entropy}. This is in contrast with WebGraph's $\zeta$ coding~\cite{boldi2004webgraph2}. 

Secondly, Zuckerli splits the nodes of~$G$ into \emph{chunks} of size $C$, where the first chunk contains the first $C$ nodes in $G$, the second chunk contains the following $C$ nodes in $G$, and so on. When list decompression is not required, we set $C=\infty$. Inside each chunk, \emph{degrees} of the nodes are stored.
Notably, the representations of node degrees requires a significant amount of bits. To improve compression, Zuckerli represents it via delta encoding, i.e. as the difference between the current degree and the previous one. As this procedure
may produce negative numbers, deltas are represented using the transformation
described in Equation~\ref{eqn:packsigned}. Delta encoding across multiple adjacency lists is of course hostile to allowing access to any adjacency lists without decoding the rest of the graph first. For this reason, Zuckerli adopts chunks.

Thirdly, while Zuckerli uses \emph{reference} lists and blocks in the same way as WebGraph (points~\ref{item:reference} and \ref{item:blocks}), the choice of the reference list and reference chain is more sophisticated. We defer its description to Section~\ref{sub:approx}.

Fourthly, Zuckerli does not use \emph{intervals}, in contrast with WebGraph (point~\ref{item:intervals}). As a form of simplification, the special representation for intervals is replaced with run-length encoding~\cite{rle67} of zero
gaps. When reading residuals, as soon as a sequence of exactly $L'$ zero gaps is read, for a global parameter $L'$, another integer is read to represent the subsequent number of zero gaps, which are not otherwise represented in the compressed representation. Since ANS does not require an integer number of bits per
symbol, and allows for very efficient representations of sequences of zeros, we set
$L'=\infty$ if list decompression is not supported.

Finally, Zuckerli modifies the representation of the \emph{residuals}, which are stored via delta encoding. The
representation chosen by WebGraph (point~\ref{item:residuals}) does not exploit the fact that an edge might already be
represented by block copies (or intervals).
For example, consider the case in which an
adjacency list contains edges $\{1, 2, 3, 4, 8, 9\}$, and edges $\{1, 2, 4\}$ are already represented by block copies.
Residuals would then be $\{3, 8, 9\}$ and the second residual would be represented by WebGraph using a delta of $4=8-3-1$.
However, this representation does not take into account the fact that not all possible delta values smaller or equal to $4$
are useful. In this example, reading a delta of $0$ from the compressed file would result in an edge value of $4$,
which would be either invalid or superfluous, as this edge is already represented through blocks.
Thus, Zuckerli modifies the delta encoding of residuals by subtracting the number of 
edges that are between the previous and the current residual edge and that are already known to be encoded using blocks. In this case, residual
edge $8$ would be represented as $3$ instead of $4$, as there are only $3$ possible edges between
$3$ and $8$ that are not already represented in the block copies.

A full example description of how Zuckerli would represent an adjacency list is shown in Figure~\ref{fig:residual}.

\begin{figure}[t]
    \centering
    \scalebox{0.9}{\begin{tikzpicture}[node distance=0.7cm]
        \node (a0) {$1$};
        \node (al) [left of=a0, node distance=1.5cm] {ref. node ($6$)};
        \node (a1) [right of=a0] {$2$};
        \node (a2) [right of=a1] {$4$};
        \node (a3) [right of=a2] {$5$};
        \node (a4) [right of=a3] {$7$};
        \node (a5) [right of=a4] {$10$};
        \node (a6) [right of=a5] {$11$};
        \node (a7) [right of=a6] {$12$};
        \node (b0) [below of=a0] {$1$};
        \node (bl) [left of=b0, node distance=1.5cm] {current node ($7$)};
        \node (b1) [right of=b0] {$2$};
        \node (b2) [right of=b1] {$3$};
        \node (b3) [right of=b2] {$4$};
        \node (b4) [right of=b3] {$8$};
        \node (b5) [right of=b4] {$9$};
        \node (b6) [right of=b5] {$10$};
        \node (b7) [right of=b6] {$11$};
        \node (b8) [right of=b7] {$12$};
        \node (b9) [right of=b8] {$13$};
        \node (c0) [below of=b0] {$3$};
        \node (cl) [left of=c0, node distance=1.5cm] {block lengths};
        \node (c1) [right of=c0] {$2$};
        \node (c2) [right of=c1] {$3$};
        \node (d0) [below of=c0] {$3$};
        \node (dl) [left of=d0, node distance=1.5cm] {block encoding};
        \node (d1) [right of=d0] {$1$};
        \node (e0) [below of=d0] {$3$};
        \node (el) [left of=e0, node distance=1.5cm] {residuals};
        \node (e1) [right of=e0] {$8$};
        \node (e2) [right of=e1] {$9$};
        \node (e3) [right of=e2] {$13$};
        \node (f0) [below of=e0] {$-4$};
        \node (fl) [left of=f0, node distance=1.5cm] {residuals delta};
        \node (f1) [right of=f0] {$3$};
        \node (f2) [right of=f1] {$0$};
        \node (f3) [right of=f2] {$0$};
        \node (g0) [below of=f0] {$2$};
        \node (gl) [left of=g0, node distance=1.5cm] {list repr.};
        \node (g1) [right of=g0] {$1$};
        \node (g2) [right of=g1] {$2$};
        \node (g3) [right of=g2] {$3$};
        \node (g4) [right of=g3] {$1$};
        \node (g5) [right of=g4] {$-4$};
        \node (g6) [right of=g5] {$3$};
        \node (g7) [right of=g6] {$0$};
        \node (g8) [right of=g7] {$0$};
        \begin{pgfonlayer}{bg}
        \fill[blue!30] ($(a0.north west)$)  rectangle ($(a2.south east)$);
        \fill[blue!30] ($(a5.north west)$)  rectangle ($(a7.south east)$);
        \fill[blue!30] ($(b0.north west)$)  rectangle ($(b1.south east)$);
        \fill[blue!30] ($(b3.north west)$)  rectangle ($(b3.south east)$);
        \fill[blue!30] ($(b6.north west)$)  rectangle ($(b8.south east)$);
        \fill[red!20] ($(b2.north west)$)  rectangle ($(b2.south east)$);
        \fill[red!20] ($(b4.north west)$)  rectangle ($(b5.south east)$);
        \fill[red!20] ($(b9.north west)$)  rectangle ($(b9.south east)$);
        \fill[red!20] ($(e0.north west)$)  rectangle ($(e3.south east)$);
        \draw[thick] ($(g0.north west)$)  rectangle ($(g0.south east)$);
        \draw[thick] ($(g1.north west)$)  rectangle ($(g1.south east)$);
        \draw[thick] ($(g2.north west)$)  rectangle ($(g2.south east)$);
        \draw[thick] ($(g3.north west)$)  rectangle ($(g4.south east)$);
        \draw[thick] ($(g5.north west)$)  rectangle ($(g8.south east)$);
        \end{pgfonlayer}

    \end{tikzpicture}
    }
    \caption{Example encoding of an adjacency list. We are encoding the adjacency list of node $7$ using the adjacency list of node $6$ as a reference. Highlighted in blue are the edges that the two nodes have in common, i.e. the blocks to be copied from the reference node adjacency list. The block encoding is performed as described in Section~\ref{sec:webgraph} (point~\ref{item:blocks}). Highlighted in red are the residual values, which are stored as follows: the first residual is encoded as the delta between the current node and itself, while the next values are encoded as $d-1$, where $d$ is the value to add to the previous residual, implicitly skipping any possible edges that have already been added though blocks. The boxes in the final list representation show, in order, the data that gets encoded: the delta of the degree of the current node with respect to the previous node, the delta (in absolute value) of the reference node with respect to the current node, the number of blocks, the block encoding, the residual deltas.}
    \label{fig:residual}
\end{figure}

\subsection{Context management}
\label{sub:ctx} 

As mentioned in Section~\ref{sec:entropy}, Zuckerli uses Huffman coding and ANS with multiple contexts, i.e. distinct probability distributions. To the best of our knowledge, while this is a well-known encoding technique, its application to graph compression is new. Here we detail how symbols are split among the different contexts.

Inside each chunk, the symbol that represents the delta-coded degree with respect to the previous
node is used to choose the distribution for the current node. Similarly,
inside a chunk, the reference number used for the last list is used to choose a
distribution for the current one.

When compressing blocks, a separate distribution is used for the first block,
all the even blocks, and all the odd blocks. This is because the first block is the only one
for which its length does not get reduced by $1$, and we expect the number of edges to be copied (odd blocks) to have a
different distribution from the number of edges to be skipped (even blocks), depending on the graph.

For delta-encoding the first residual with respect to the current node, the
symbol that would be used to represent the number of residuals defines which
distribution to use. This is because a list with a high number of residuals will 
likely be harder to predict. 

Finally, for all other residual deltas, the symbol that was used to encode the
previous one is used to choose the corresponding probability distribution for the
current delta.

We remark that each probability distribution used by Zuckerli is stored in the compressed file, and is not changed as edges are decoded.

\subsection{Choice of reference list and chain}
\label{sub:approx}

We explain how Zuckerli selects reference lists to be used during compression. 
As previously discussed, we may either represent a node's list explicitly or, if we use a reference, 
we represent the difference from the list of its reference.

% This section explains the algorithms that the encoder uses to select the best reference list to be used during compression.

To make an effective choice, we need to estimate the amount of bits that the algorithm will use to compress an adjacency list using a given reference.
Since we use an adaptive entropy model, this is not a simple task, as choices for one list might affect
probabilities for all other ones. 

We choose to use an iterative approach previously used by Zopfli~\cite{alakuijala2013data}, a general compression algorithm.
We initialize symbol probabilities with a
simple fixed model (all symbols have equal probability), and then choose reference
lists assuming these will be the final costs. We then update the symbol
probabilities given by the chosen reference lists and repeat the procedure with
the new probability distribution. This process is then repeated a constant
number of times. 

We now consider the two types of compression separately:

\medskip
\noindent\textbf{Full decompression.} 
In this case, there is no limitation on the length of the reference chain used
by a single node, i.e., a reference node may itself have a reference node, and so on; we obtain an optimal solution with the greedy strategy, choosing the reference node that gives the best compression out of all the ones available in the window of the current node, i.e., the $W$ preceding nodes.

\medskip
\noindent\textbf{List decompression.} 
To decompress a single list, we must also decompress its reference chain: when access to single lists is requested, more care is required to
select good references while avoiding reference chains longer than a given threshold $R$.\footnote{Each node $u$ may refer in turn to any of its $W$ preceding ones during a hop, which makes $R$ unrelated to $W$: indeed, $R$ is the maximum number of these hops.}

\medskip
For example, imagine these are the lists of nodes $1$,$2$ and $3$:
\[
1: \{3,4,7\},~ 2 : \{3,4,7,9\},~ 3 : \{4,7,9\} 
\]

We may want to represent $2$'s list using $1$'s as a reference: this way we do not need to represent $3$, $4$, and $7$, but just the node $9$ in the difference; similarly, if we represent $3$'s list using $2$'s as reference, we just need to omit node $3$. However, in order to decompress $3$'s list we will need to read (hence decompress) the list of its reference $2$, which in turn requires decompressing $1$'s list. The longer the chain, the longer the decompression time: the parameter $R$ allows us to keep this overhead under control.

\medskip

We can formally state the problem of choosing the references as follows. We are given a directed acyclic graph $D$, where the nodes represents the adjacency lists. There is an arc between two nodes if one adjacency list can refer to the other. The weight of the arc corresponds to the number of bits saved by choosing that reference. The larger the weights, the better the compression gain. Thus, we aim at finding a maximum-weight directed forest $O$ for $D$, where each node has out-degree at most one (its reference), and there are no directed paths longer than $R$ (i.e.~a reference chain longer than~$R$). Finding an optimal solution seems not trivial, and it is unclear whether it can be done in polynomial-time.~\footnote{We speculate it may be NP-complete due to similarities with maximum directed cuts~\cite{papadimitriou1991optimization}.}

%To solve this, WebGraph proceeds greedily as above, but ignores all lists in the window that would produce reference chain longer than $R$, without changing decisions taken in previous nodes.
% A simple solution (which is used by WebGraph) is to discard all lists in the window that would produce a too long reference chain, without changing decisions taken in previous nodes.

Zuckerli uses an efficient heuristic with approximation guarantees. 
Given $D$, it first builds the optimal directed forest $F$, ignoring the constraint that directed paths cannot be longer than $R$ (this corresponds to the solution of the full decompression case).
%
% each tree arc is weighted with the number of bits that are saved by using the
% parent node as a reference for the child node (such an optimal tree can easily
% be constructed by the greedy algorithm that is used when no access to single
% lists is required). 

Instead of solving our problem on $D$ as we formulated above, Zuckerli computes an optimal sub-forest $H$ on $F$, as the latter be found by the following dynamic programming algorithm,
%----
%solves a dynamic programming problem on $T$, 
answering the question \textsl{``what is the sub-forest $H$ of maximum
weight that is contained in the current subforest of $F$ and does not have paths of length $R+1$?''}.% vedi commento
%while performing a post-order traversal of $T$.
%---

Clearly, $H$ is not necessarily the optimal solution for $D$, as it is computed for its subgraph $F$. However, there may still be arcs of $D$ that were not in $F$, but can now be added to $H$ without creating long chains. Zuckerli tries to extend $H$ with such arcs in a greedy way, obtaining the final heuristic solution.

\medskip
\noindent\textbf{Approximation guarantee.}
Interestingly, our heuristics not only works quite well in practice, but it also provides a guaranteed $(1-\frac{1}{R+1})$-approximation of the optimal solution on $D$, i.e. of the maximum number of bits to be saved. 

To see why, let $O$ be the optimal solution, and let $w_O$, $w_F$ and $w_H$ be the total weights of $O$, $F$, and $H$, respectively.

Next, let $H'$ be a sub-forest of $F$ obtained by splitting the arcs of $F$ in $R+1$ groups, depending on their distance from the root of their tree in $F$ modulo $R+1$, then removing the group of smallest weight; it is evident that $H'$ has no paths longer than $R$, and that its weight $w_{H'}$ is at least $(1-\frac{1}{R+1}) w_F$, as the weight of smallest of the $R+1$ groups could not be more than $\frac{1}{R+1} w_F$.
Now observe the following:
% of the optimal sub-forest extracted by the dynamic programming algorithm, and the weight $w_O$ of the forest that represents the best possible choice of reference nodes, we have:

% if we consider the total weight $w_T$ of $T$, the weight $w_H$ of the optimal sub-forest extracted by the dynamic programming algorithm, and the weight $w_O$ of the forest that represents the best possible choice of reference nodes, we have:

\begin{itemize}
  \item $w_F \ge w_O$, as $F$ is the optimal solution for $R = \infty$.% a problem with less constraints;
  \item $w_H \ge w_{H'} \ge (1-\frac{1}{R+1})w_F$, as $H'$ is a sub-forest of $F$, and $H$ contains the optimal sub-forest of $F$ (both with path length bounded by $R$).
%   In particular, the forest obtained by erasing the set of arcs of minimum total weight will have a weight of at least $(1-\frac{1}{R+1})w_T$. As $w_H$ is the optimal forest that satisfies the maximum path length constraint, its weight will be at least as large;
  \item Thus, $w_H \ge (1-\frac{1}{R+1})w_F \ge (1-\frac{1}{R+1})w_O$, which proves the approximation bound.
\end{itemize}

\medskip
\noindent \textbf{Details on computing the optimal sub-forest of \boldmath $F$.}
Given a sub-forest $F'$ of $F$ rooted in the node $x$, let $M_i(x)$ be the maximum weight sub-forest of $F'$ that has no paths longer than $R$, and in which the root $x$ is in no path longer than $i$. If $r_j$ are the roots of $F$, $\bigcup_j M_R(r_j)$ is the optimal sub-forest of $F$ we are looking for.
We implement a dynamic programming procedure based on the following invariant: if, for all sub-forests rooted in each child $y$ of $x$, we know $M_i(y)$ for each $i\in\{0,\ldots, R\}$, we can compute $M_i(x)$ for each $i\in\{0,\ldots, R\}$.

\begin{table}[t]
  \centering
  \begin{tabular}{lrrr}
    \toprule
    name & nodes & edges \\
    \midrule
    \texttt{cnr-2000} & $\num{325557}$ & $\num{3216152}$ \\
    \texttt{in-2004} & $\num{1382908}$ & $\num{16917053}$ \\
    \texttt{eu-2005} & $\num{862664}$ & $\num{19235140}$ \\
    \texttt{hw-2009} & $\num{1139905}$ & $\num{113891327}$ \\
    \texttt{uk-2002} & $\num{18520486}$ & $\num{298113762}$ \\
    \texttt{tw-2010} & $\num{41652230}$ & $\num{1468365182}$ \\
    \texttt{uk-2007-02} & $\num{110123614}$ & $\num{3944932566}$ \\
    \texttt{eu-2015} & $\num{1070557254}$ & $\num{91792261600}$ \\
    \bottomrule
  \end{tabular}
  \caption{Graphs used during experiments, with node and edge counts. All graphs are web graphs, except \texttt{hw-2009} (\texttt{hollywood-2009}) and \texttt{tw-2010} (\texttt{twitter-2010}), which are social networks.\label{tab:sizes}}
\end{table}
% twocolumn <3
\begin{table*}[ht] % updated
  \centering
  \begin{tabular}{lrrrrrrr}
    \toprule
    && \multicolumn{6}{c}{Size (bits per edge)} \\
    & $k$ & $3$ & $4$ & $4$ & $4$ & $5$ & $5$ \\
    & $i$ & $1$ & $1$ & $2$ & $2$ & $2$ & $2$ \\
    & $j$ & $0$ & $0$ & $0$ & $1$ & $0$ & $1$ \\
    \midrule
    \texttt{cnr-2000-hc} && $\mathbf{1.86}$ & $\num{1.87}$ & $\num{1.88}$ & $\num{1.89}$ & $\num{1.90}$ & $\num{1.91}$ \\
    \texttt{cnr-2000} && $\mathbf{2.23}$ & $\num{2.24}$ & $\num{2.25}$ & $\num{2.29}$ & $\num{2.26}$ & $\num{2.31}$ \\
    \texttt{in-2004-hc} && $\mathbf{1.32}$ & $\num{1.33}$ & $\num{1.33}$ & $\num{1.33}$ & $\num{1.33}$ & $\num{1.34}$ \\
    \texttt{in-2004} && $\mathbf{1.69}$ & $\mathbf{1.69}$ & $\num{1.71}$ & $\num{1.77}$ & $\num{1.72}$ & $\num{1.79}$ \\
    \texttt{eu-2005-hc} && $\num{2.49}$ & $\num{2.49}$ & $\mathbf{2.47}$ & $\mathbf{2.47}$ & $\mathbf{2.47}$ & $\mathbf{2.47}$ \\
    \texttt{eu-2005} && $\mathbf{2.88}$ & $\num{2.89}$ & $\mathbf{2.88}$ & $\num{2.92}$ & $\mathbf{2.88}$ & $\num{2.93}$ \\
    \texttt{uk-2002-hc} && $\num{1.38}$ & $\num{1.38}$ & $\mathbf{1.37}$ & $\mathbf{1.37}$ & $\num{1.38}$ & $\mathbf{1.37}$ \\
    \texttt{uk-2002} && $\mathbf{1.75}$ & $\num{1.76}$ & $\num{1.78}$ & $\num{1.87}$ & $\num{1.79}$ & $\num{1.89}$ \\
    \texttt{tw-2010-hc} && $\mathbf{11.99}$ & $\num{12.00}$ & $\mathbf{11.99}$ & $\mathbf{11.99}$ & $\num{12.00}$ & $\num{12.00}$ \\
    \texttt{tw-2010} && $\mathbf{12.12}$ & $\num{12.13}$ & $\mathbf{12.12}$ & $\num{12.58}$ & $\num{12.26}$ & $\num{12.62}$ \\
    \texttt{uk-2007-02-hc} && $\num{0.92}$ & $\num{0.92}$ & $\num{0.92}$ & $\mathbf{0.91}$ & $\num{0.92}$ & $\num{0.93}$ \\
    \texttt{uk-2007-02} && $\mathbf{1.20}$ & $\mathbf{1.20}$ & $\num{1.22}$ & $\num{1.30}$ & $\num{1.23}$ & $\num{1.31}$ \\
    \bottomrule
  \end{tabular}
  \caption{Effects of changing hybrid integer encoding parameters.\label{tab:hyb}}
\end{table*}

First, as paths are always directed from nodes to their parent, observe that we can consider each child $y$ of $x$ independently. 
Furthermore, $M_i(x)$ is made as follows: if the arc $(x,y)$ is taken, then $y$ in its sub-forest may only partake in paths of length at most $i-1$; on the other hand, if we do not choose $(x,y)$, $y$ may partake in paths of any length up to $R$. Finally, for the base case, observe that for any leaf $l$ of $F$, $M_i(l) = \emptyset$.
We thus obtain each $M_i(x)$ by the following formula:

$$M_i(x) = \bigcup_{y\in \children(x)}\maxw \left( M_R(y),  \{(x,y)\} \cup M_{i-1}(y) \right)$$

where $\children(x)$ are the children of $x$ in $F$, and $\maxw(A,B)$ returns the set of arcs having greater weight between $A$ and $B$ (breaking ties arbitrarily). 

Finally, we give a brief remark on the complexity. This is important since a trivial implementation would take quadratic time and space to represent each set $M_i()$, making this approach unfeasible on graphs with millions of nodes.
However, we can implement it in $O(nR)$ time and space, where $n$ is the number of nodes in $F$, as follows. We can first run the above dynamic programming algorithm, but associate with each $M_i(y)$ just its weight. Furthermore, we keep track for each $M_i(x)$ of which was the choice performed on each child $y$ of $x$ (i.e., whether we used $(x,y)$ or not). Computing the weights of $M_i(x)$ this way takes just $O(1)$ time for each child, costing us in total $O(nR)$ as $F$ as $O(n)$ arcs. With this information, we can reconstruct exactly which arcs are used in the optimal solution $M_R(r)$ in a top-down manner by looking at the information about its children we previously computed.

\section{Experiments}
\label{sec:experiments} 

In order to evaluate the efficiency of Zuckerli, we first study the effects of various choices of parameters on compressed size.
We also evaluate the effectiveness of the approximation algorithm for reference selection.

We then compare the compression ratio of Zuckerli with respect to existing state-of-the-art compression systems for
large graphs, either with novel experiments (WebGraph~\cite{boldi2004webgraph}, Graph Compression by BFS\cite{apostolico2009graph}) or by referring to the experiments in the relevant papers (LogGraph~\cite{besta2018log}, $k^2$-tree~\cite{brisaboa2009k} and $2$D-Block Trees~\cite{brisaboa2018two}). We remark that the proposed scheme does not change the order of nodes before compression, and as such a comparison with works that propose algorithms to find a better node permutation (such as~\cite{dhulipala2016compressing}) is out of scope of this experimental comparison, although it is an interesting direction for future work.

To evaluate the CPU and memory usage of Zuckerli, we compare its decompression time and memory usage with the corresponding metrics for WebGraph. 
Moreover, we compare the running time of a depth-first search and a breadth-first search on Zuckerli-compressed graphs,
on Webgraph-compressed graphs and on uncompressed graphs. 

Finally, to evaluate the parallelism of the code, we compute the speedup achieved by Zuckerli on an edge-summing problem when running on $2$, $4$, $8$, $16$, $32$ and $64$ cores.

For all experiments where list decompression is required, $R$ is set to $3$ (similarly to the compressed WebGraph files that used for comparisons), the chunk size $C$ is set to $32$, and the minimum run of $0$s to use RLE $L'$ is set to $3$.

The code to run the experiments was written in C++ and compiled with \texttt{clang++-10}; it is available at \url{https://github.com/google/zuckerli}.
The experiments were ran on a $32$-core AMD 3970X CPU (with hyperthreading) with $256$GB of RAM.

\subsection{Datasets}

To run the comparisons, we use graphs from the WebGraph
corpus~\cite{boldi2004webgraph,BRSLLP,BMSB}, which are available at
\url{http://law.di.unimi.it/datasets.php}. The datasets we use include both social
networks and web graphs, with a number of edges varying from a few millions to
$91$ billions and a number of nodes varying from a few hundred thousands to $1$
billion. More details about the graphs can be found in Table~\ref{tab:sizes}.
When reporting results, graphs with a \texttt{-hc} suffix represent the full
decompression versions, while other graphs represent the compressed versions also supporting list decompression.

\subsection{Parameter Choice}
We first investigate the effect of the parameters controlling the integer
encoding scheme, trying different combinations
of the number of bits that are included in the entropy-coded part
and the number of integers that are entropy coded as-is. The results are shown in
Table~\ref{tab:hyb}. They show that using more fine-grained integer
representations, i.e. entropy-coding more bits or having more direct-coded
integers, does not give significant improvements or even worsens the compression ratio.

\begin{table}[ht] % updated
  \centering
  \begin{tabular}{lrrr}
    \toprule
    name & \multicolumn{3}{c}{Size (bits per edge)} \\
    & $W=16$ & $W=32$ & $W=64$  \\
    \midrule
    \texttt{cnr-2000-hc} & $\num{1.95}$ & $\num{1.87}$ & $\mathbf{1.82}$ \\
    \texttt{cnr-2000} & $\num{2.31}$ & $\num{2.24}$ & $\mathbf{2.20}$ \\
    \texttt{in-2004-hc} & $\num{1.34}$ & $\num{1.33}$ & $\mathbf{1.31}$ \\
    \texttt{in-2004}    & $\num{1.71}$ & $\num{1.69}$ & $\mathbf{1.68}$ \\
    \texttt{eu-2005-hc} & $\num{2.60}$ & $\num{2.49}$ & $\mathbf{2.43}$ \\
    \texttt{eu-2005} & $\num{2.99}$ & $\num{2.89}$ & $\mathbf{2.83}$\\
    \texttt{uk-2002-hc} & $\num{1.42}$ & $\num{1.38}$ & $\mathbf{1.35}$ \\
    \texttt{uk-2002} & $\num{1.79}$ & $\num{1.76}$ & $\mathbf{1.73}$\\
    \texttt{tw-2010-hc}& $\num{12.05}$ & $\num{12.00}$ & $\mathbf{11.95}$ \\
    \texttt{tw-2010}   & $\num{12.18}$ & $\num{12.13}$ & $\mathbf{12.09}$ \\
    \texttt{uk-2007-02-hc} & $\num{0.95}$ & $\num{0.92}$ & $\mathbf{0.90}$ \\
    \texttt{uk-2007-02} & $\num{1.23}$ & $\num{1.20}$ & $\mathbf{1.18}$ \\
    \bottomrule
  \end{tabular}
  \caption{Effects of changing window size.\label{tab:wnd}}
\end{table}

Next, we compare the effect of changing the window size $W$, choosing between values of
$16$, $32$, and $64$. The results are reported in Table~\ref{tab:wnd}. They show that
increasing window size gives significant, although diminishing, savings on compressed
size.

\begin{table}[t]
  \centering
  \begin{tabular}{lrrr} %updated
    \toprule
    name & \multicolumn{3}{c}{Size (bits per edge)} \\
    & $1$ iter. & $2$ iter. & $3$ iter.  \\
    \midrule
    \texttt{cnr-2000-hc} & $\num{1.87}$ & $\mathbf{1.84}$ & $\mathbf{1.84}$ \\
    \texttt{cnr-2000}    & $\num{2.24}$ & $\mathbf{2.19}$ & $\mathbf{2.19}$ \\
    \texttt{in-2004-hc} & $\num{1.33}$ & $\mathbf{1.31}$ & $\mathbf{1.31}$ \\
    \texttt{in-2004}    & $\num{1.69}$ & $\mathbf{1.65}$ & $\mathbf{1.65}$ \\
    \texttt{eu-2005-hc} & $\num{2.49}$ & $\mathbf{2.46}$ & $\mathbf{2.46}$ \\
    \texttt{eu-2005}    & $\num{2.89}$ & $\mathbf{2.83}$ & $\mathbf{2.83}$\\
    \texttt{uk-2002-hc} & $\num{1.38}$ & $\mathbf{1.36}$ & $\mathbf{1.36}$ \\
    \texttt{uk-2002} & $\num{1.76}$ & $\mathbf{1.72}$ & $\mathbf{1.72}$\\
    \texttt{tw-2010-hc}& $\num{12.00}$ & $\mathbf{11.97}$ & $\mathbf{11.97}$ \\
    \texttt{tw-2010}   & $\num{12.13}$ & $\mathbf{12.12}$ & $\mathbf{12.12}$ \\
    \texttt{uk-2007-02-hc} & $\num{0.92}$ & $\mathbf{0.91}$ & $\mathbf{0.91}$ \\
    \texttt{uk-2007-02} & $\num{1.20}$ & $\mathbf{1.18}$ & $\mathbf{1.18}$ \\
    \bottomrule
  \end{tabular}
  \caption{Effects of changing number of iterations for reference list selection.\label{tab:zopf}}
\end{table}

Finally, we compare the effect of changing the number of iterations through which reference
lists are chosen (see Section~\ref{sub:approx}), varying between $1$ (corresponding to only using the simple fixed model) to
$3$. The results are shown in Table~\ref{tab:zopf}. They show that using a non-fixed model provides
significant savings compared to the fixed one. On the other hand, further refinement of this model
does not improve the compressed size, and is thus not worth the extra encoding effort.

As a consequence of these results, we perform further experiments using $k=4$, $i=1$, $j=0$, $W=32$,
and $2$ rounds of reference selection. We remark that $W=64$ would have achieved better
compression, but the WebGraph dataset was compressed using $W=32$. We therefore pick this value for ease
of comparison.

\begin{table}[t] %updated
  \centering
  \begin{tabular}{lrr}
    \toprule
    name & \multicolumn{2}{c}{bits/edge} \\
     & greedy & approx \\
    \midrule
    \texttt{cnr-2000} & $\num{2.49}$ & $\mathbf{2.24}$ \\
    \texttt{in-2004} & $\num{1.82}$ & $\mathbf{1.69}$ \\
    \texttt{eu-2005} & $\num{3.18}$ & $\mathbf{2.89}$ \\
    \texttt{uk-2002} & $\num{1.95}$ & $\mathbf{1.76}$ \\
    \texttt{tw-2010} & $\num{12.29}$ & $\mathbf{12.13}$ \\
    \texttt{uk-2007-02} & $\num{1.36}$ & $\mathbf{1.20}$ \\
    \bottomrule
  \end{tabular}
  \caption{Comparison of the compressed size achieved by using the greedy algorithm used by WebGraph for reference selection and the size achieved by our approximation algorithm described in Section~\ref{sub:approx}.\label{tab:greedy}}
\end{table}

\begin{table}[t] %updated
  \centering
  \begin{tabular}{lrr}
    \toprule
    name & \multicolumn{2}{c}{bits/edge} \\
     & no ctx model & default \\
    \midrule
    \texttt{cnr-2000-hc} & $\num{2.17}$ & $\mathbf{1.84}$ \\
    \texttt{cnr-2000} & $\num{2.47}$ & $\mathbf{2.19}$ \\
    \texttt{in-2004-hc} & $\num{1.55}$ & $\mathbf{1.31}$ \\
    \texttt{in-2004} & $\num{1.86}$ & $\mathbf{1.65}$ \\
    \texttt{eu-2005-hc} & $\num{2.84}$ & $\mathbf{2.46}$ \\
    \texttt{eu-2005} & $\num{3.14}$ & $\mathbf{2.83}$ \\
    \texttt{uk-2002-hc} & $\num{1.58}$ & $\mathbf{1.36}$ \\
    \texttt{uk-2002} & $\num{1.92}$ & $\mathbf{1.72}$ \\
    \texttt{tw-2010-hc} & $\num{13.21}$ & $\mathbf{11.97}$ \\
    \texttt{tw-2010} & $\num{13.27}$ & $\mathbf{12.12}$ \\
    \texttt{uk-2007-02-hc} & $\num{1.04}$ & $\mathbf{0.91}$ \\
    \texttt{uk-2007-02} & $\num{1.31}$ & $\mathbf{1.20}$ \\
    \bottomrule
  \end{tabular}
  \caption{Effects of disabling Zuckerli's context model.\label{tab:ctx}}
\end{table}

% twocolumn <3
\begin{table*}[t]
  \centering
  \begin{tabular}{lcrrrrrr} %updated
    \toprule
    name & \multicolumn{1}{c}{compression} & \multicolumn{5}{c}{bits/edge} \\
         &   \multicolumn{1}{c}{speed ($10^6 e/s$)}     & Zuckerli & \multicolumn{2}{c}{WebGraph} & \multicolumn{2}{c}{GCBFS} \\
    \midrule
    \texttt{cnr-2000-hc}    & $\num{1.01}$ & $\num{1.84}$ & $\num{2.45}$ & $75\%$ & $\num{1.88}$ & $98\%$\\
    \texttt{cnr-2000}       & $\num{0.89}$ & $\num{2.19}$ & $\num{3.12}$ & $71\%$ & $\num{2.72}$ & $80\%$ \\
    \texttt{in-2004-hc}     & $\num{1.19}$ & $\num{1.31}$ & $\num{1.76}$ & $75\%$ & $\num{1.42}$ & $92\%$ \\
    \texttt{in-2004}        & $\num{1.03}$ & $\num{1.65}$ & $\num{2.15}$ & $77\%$ & $\num{2.16}$ & $76\%$ \\
    \texttt{eu-2005-hc}     & $\num{1.03}$ & $\num{2.46}$ & $\num{3.16}$ & $78\% $ & $\num{2.81}$ & $87\%$\\
    \texttt{eu-2005}        & $\num{0.97}$ & $\num{2.83}$ & $\num{3.72}$ & $76\% $ & $\num{3.50}$ & $80\%$\\
    \texttt{uk-2002-hc}     & $\num{1.15}$ & $\num{1.36}$ & $\num{1.80}$ & $76\% $ & $\num{1.71}$ & $79\%$\\
    \texttt{uk-2002}        & $\num{1.03}$ & $\num{1.72}$ & $\num{2.24}$ & $77\% $ & $\num{2.43}$ & $70\%$\\
    \texttt{hw-2009-hc}& $\num{0.69}$ & $\num{4.26}$ & $\num{4.80}$ & $89\% $ & $\num{7.29}$ & $58\%$ \\
    \texttt{hw-2009}   & $\num{0.60}$ & $\num{4.47}$ & $\num{4.94}$ & $90\% $ & $\num{7.51}$ & $59\%$\\
    \texttt{tw-2010-hc}& $\num{0.50}$ & $\num{11.97}$ & $\num{13.89}$ & $86\% $ & $\num{15.34}$ & $78\%$ \\
    \texttt{tw-2010}   & $\num{0.42}$ & $\num{12.12}$ & $\num{14.46}$ & $84\% $ & $\num{15.21}$ & $80\%$\\
    \texttt{uk-2007-02-hc}  & $\num{1.63}$ & $\num{0.91}$ & $\num{1.18}$ & $77\% $ & $\num{1.28}$ & $71\%$\\
    \texttt{uk-2007-02}     & $\num{1.53}$ & $\num{1.18}$ & $\num{1.56}$ & $75\% $ & $\num{1.80}$ & $65\%$\\
    \texttt{eu-2015-hc}     & $\num{1.46}$ & $\num{0.74}$ & $\num{0.89}$ & $82\% $ & \multicolumn{2}{c}{-}\\
    \texttt{eu-2015}        & $\num{1.34}$ & $\num{0.92}$ & $\num{1.20}$ & $77\% $ & \multicolumn{2}{c}{-}\\
    \bottomrule
  \end{tabular}
  \caption{Comparison of compressed size between Zuckerli, WebGraph and Graph Compression with BFS, with compression speed for Zuckerli. The GCBFS encoder crashed when compressing \texttt{eu-2015}.\label{tab:webgraph}}
\end{table*}

\subsection{Effect of Approximation Algorithm and Context Modeling}

We evaluate the gain from using the improved algorithm for reference selection (in Section~\ref{sub:approx}), as opposed to the simple greedy algorithm used by WebGraph. The results are shown in Table~\ref{tab:greedy}. We remark that, as the reference selection is employed only when list decompression is supported, the table does not report results for the \texttt{-hc} version of the graphs.

We also report the effects of disabling Zuckerli's context model, by using the same probability distribution for all the entropy coded symbols. The results are shown in Table~\ref{tab:ctx}.

The results show that the gains from the approximation algorithm are significant, reaching up to $12\%$
for web graphs, and also providing some benefits for social networks like \texttt{tw-2010}. The gains from the context model are similar.

We remark that this improvement is significant in a lossless compression context. In comparison, one of the most well-known advances in general purpose compression, the Burrows-Wheeler Transform~\cite{burrows1994block}, achieved roughly a $16\%$ size reduction compared to previous approaches.

\subsection{Compression Results and Resource Usage}
For the chosen set of parameters, we report the compression speed and the resulting compression ratio on various graphs.
We also compare the resulting compressed size with the ones
achieved by WebGraph and by Graph Compression By BFS (GCBFS). To perform this comparison, we use the files available from the WebGraph
corpus itself, without any recompression, and the implementation of GCBFS that is publicly available, with parameters $l=10000$ for full decompression and $l=8$ for list decompression. The results
are shown in Table~\ref{tab:webgraph}. They show that Zuckerli typically achieves $20\%$ to
$30\%$ size savings when compared to WebGraph on web graphs, and $10\%$ to $15\%$ size savings on social networks. In comparison, GCBFS achieves worse compression ratios than WebGraph in the larger datasets (\texttt{hw-2009}, \texttt{tw-2010}, \texttt{uk-2007}), and worse compression ratios than Zuckerli in all datasets (by up to $42\%$). Moreover, the decompression speed reported in the original paper is comparable with the one of WebGraph. Thus, we decide to run the remaining experiments comparing only with WebGraph.

We also compare Zuckerli's compression ratios to those achieved by $k^2$-trees~\cite{brisaboa2009k} and
$2$D-Block Trees~\cite{brisaboa2018two}. While those data structures allow for single edge
queries, Zuckerli only allows, in its least dense configurations, for individual adjacency
list queries. Thus, the methods are not directly comparable. However, according to the results
reported in~\cite{brisaboa2018two}, both representations are significantly less dense than
Zuckerli, with the best of the two producing compressed representations bigger by $30\%$ or more. Further,
according to the reported speed, the faster of the methods is able to process roughly $200$
thousand edges per second, due to the intense use of sophisticated succinct data structures causing many cache misses,
which is orders of magnitude slower than Zuckerli.

Finally, while we did not perform a direct comparison with LogGraph~\cite{besta2018log}, we remark that while it offers improved performance for list access compared to WebGraph, it does not achieve better compression ratios, as reported in~\cite{besta2018log} (see also Appendix~\ref{sec:apx}).

\begin{table}[t] % updated
  \centering
% \scalebox{.93}{
  \begin{tabular}{lrrrrrr}
    \toprule
    name & deg. & ref. & block & \multicolumn{2}{c}{residuals} & total \\
    & & & & first & oth. & \\
    \midrule
    \texttt{cnr-2000-hc} & $\num{0.24}$ & $\num{0.23}$ & $\num{0.36}$ & $\num{0.34}$ & $\num{0.65}$ & $\num{1.84}$ \\
    \texttt{cnr-2000} & $\num{0.27}$ & $\num{0.24}$ & $\num{0.33}$ & $\num{0.41}$ & $\num{0.92}$ & $\num{2.19}$ \\
    \texttt{in-2004-hc} & $\num{0.20}$ & $\num{0.17}$ & $\num{0.26}$ & $\num{0.24}$ & $\num{0.44}$ & $\num{1.31}$ \\
    \texttt{in-2004} & $\num{0.22}$ & $\num{0.19}$ & $\num{0.24}$ & $\num{0.32}$ & $\num{0.68}$ & $\num{1.65}$ \\
    \texttt{eu-2005-hc} & $\num{0.15}$ & $\num{0.14}$ & $\num{0.40}$ & $\num{0.32}$ & $\num{1.45}$ & $\num{2.46}$ \\
    \texttt{eu-2005} & $\num{0.16}$ & $\num{0.14}$ & $\num{0.36}$ & $\num{0.38}$ & $\num{1.79}$ & $\num{2.83}$ \\
    \texttt{uk-2002-hc} & $\num{0.19}$ & $\num{0.16}$ & $\num{0.23}$ & $\num{0.24}$ & $\num{0.54}$ & $\num{1.36}$ \\
    \texttt{uk-2002} & $\num{0.20}$ & $\num{0.16}$ & $\num{0.21}$ & $\num{0.32}$ & $\num{0.80}$ & $\num{1.72}$ \\
    \texttt{hw-2009-hc} & $\num{0.05}$ & $\num{0.02}$ & $\num{0.34}$ & $\num{0.09}$ & $\num{3.73}$ & $\num{4.26}$ \\
    \texttt{hw-2009} & $\num{0.05}$ & $\num{0.02}$ & $\num{0.33}$ & $\num{0.10}$ & $\num{3.95}$ & $\num{4.47}$ \\
    \texttt{tw-2010-hc} & $\num{0.15}$ & $\num{0.09}$ & $\num{0.30}$ & $\num{0.61}$ & $\num{10.21}$ & $\num{11.97}$ \\
    \texttt{tw-2010} & $\num{0.15}$ & $\num{0.09}$ & $\num{0.27}$ & $\num{0.62}$ & $\num{10.29}$ & $\num{12.12}$ \\
    \texttt{uk-2007-02-hc} & $\num{0.09}$ & $\num{0.07}$ & $\num{0.14}$ & $\num{0.14}$ & $\num{0.42}$ & $\num{0.91}$ \\
    \texttt{uk-2007-02} & $\num{0.09}$ & $\num{0.07}$ & $\num{0.12}$ & $\num{0.19}$ & $\num{0.60}$ & $\num{1.18}$ \\
    \bottomrule
  \end{tabular}
  \caption{Breakdown of bit allocation, reported as bits per edge.\label{tab:split}}
% }
\end{table}

\begin{figure}
\centering
\begin{tikzpicture}
\begin{axis}[xlabel={Cores}, ylabel={Speedup}, ymin=1, ymax=64, xmin=1, grid=major, xmax=64, grid style={dashed, gray!20}, xtick={1,4,8,16,64}, ytick={1,4,8,16,32,64}]
\addplot[] coordinates {(1,1)(2,1.94)(4,3.62)(8,6.73)(16,12.89)(32,20.74)(64,31.23)};
\addplot[gray, dashed] coordinates {(1,1)(64,64)};
\end{axis}
\end{tikzpicture}
\caption{Speedup obtained by Zuckerli when computing the sum of the endpoints of all
the edges using a variable number of cores on \texttt{uk-2007-02}.} \label{fig:parallelism}
\end{figure}
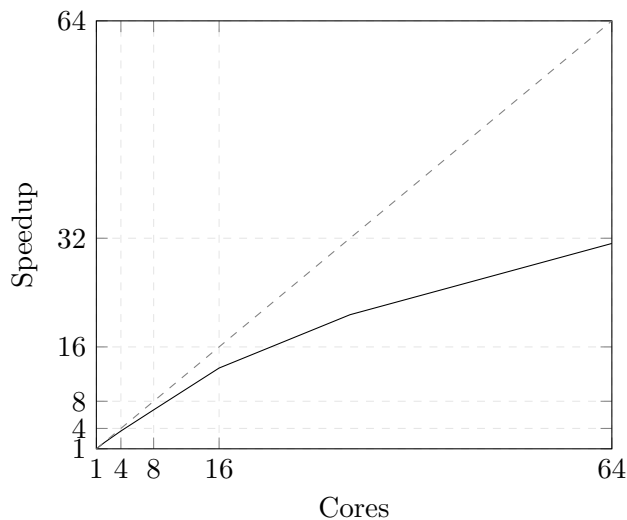

We also explore how the bit budget of Zuckerli is spent across the various parts of the graph
that get encoded: degrees, references, blocks, and residuals, with the first residual being
considered separately. The results are shown in Table~\ref{tab:split}. They show a remarkable difference
between web graphs and social networks. Indeed, in social networks, almost all the bits are spent
encoding residuals, while in web graphs the fraction of bits used for residuals is not as significant.
This can be explained by the greater effectiveness of the block copying mechanism on web graphs,
due to greater similarity in outgoing adjacency lists.

\subsection{Performance Evaluation}
\begin{table*}[t]
  \centering
  \scalebox{0.8}{
  \begin{tabular}{lrrrrrrrrr}
    \toprule
         & \multicolumn{3}{c}{uncompressed} & \multicolumn{3}{c}{Zuckerli} & \multicolumn{3}{c}{WebGraph} \\
    name & time & $\mu s/\text{adj list}$ & memory & time & $\mu s/\text{adj list}$ & memory & time & $\mu s/\text{adj list}$ & memory \\
    \midrule
    \texttt{cnr-2000} DFS & $\num{15}$ & $\num{0.04}$ & $\num{17}$ & $\num{300}$ & $\num{0.92}$ & $\num{10}$ & $\num{395}$ & $\num{1.21}$ & $\num{186}$ \\
    \texttt{cnr-2000} BFS & $\num{13}$ & $\num{0.04}$ & $\num{17}$ & $\num{302}$ & $\num{0.92}$ & $\num{10}$ & $\num{389}$ & $\num{1.19}$ & $\num{181}$ \\
    \texttt{in-2004} DFS & $\num{72}$ & $\num{0.05}$ & $\num{79}$ & $\num{1319}$ & $\num{0.95}$ & $\num{32}$ & $\num{1300}$ & $\num{0.94}$ & $\num{262}$ \\
    \texttt{in-2004} BFS & $\num{68}$ & $\num{0.05}$ & $\num{79}$ & $\num{1326}$ & $\num{0.95}$ & $\num{32}$ & $\num{1392}$ & $\num{1.00}$ & $\num{408}$ \\
    \texttt{eu-2005} DFS & $\num{89}$ & $\num{0.10}$ & $\num{84}$ & $\num{1278}$ & $\num{1.48}$ & $\num{30}$ & $\num{1542}$ & $\num{1.78}$ & $\num{293}$ \\
    \texttt{eu-2005} BFS & $\num{80}$ & $\num{0.09}$ & $\num{84}$ & $\num{1285}$ & $\num{1.49}$ & $\num{30}$ & $\num{1764}$ & $\num{2.04}$ & $\num{381}$ \\
    \texttt{uk-2002} DFS & $\num{2791}$ & $\num{0.15}$ & $\num{1315}$ & $\num{20808}$ & $\num{1.12}$ & $\num{422}$ & $\num{17974}$ & $\num{0.97}$ & $\num{1778}$ \\
    \texttt{uk-2002} BFS & $\num{1556}$ & $\num{0.08}$ & $\num{1314}$ & $\num{21256}$ & $\num{1.14}$ & $\num{431}$ & $\num{19865}$ & $\num{1.07}$ & $\num{1956}$ \\
    \texttt{hw-2009} DFS & $\num{328}$ & $\num{0.28}$ & $\num{458}$ & $\num{3922}$ & $\num{3.44}$ & $\num{147}$ & $\num{6499}$ & $\num{5.70}$ & $\num{451}$ \\
    \texttt{hw-2009} BFS & $\num{320}$ & $\num{0.28}$ & $\num{458}$ & $\num{3871}$ & $\num{3.39}$ & $\num{148}$ & $\num{6366}$ & $\num{5.58}$ & $\num{429}$ \\
    \texttt{tw-2010} DFS & $\num{11403}$ & $\num{0.27}$ & $\num{6069}$ & $\num{115120}$ & $\num{2.76}$ & $\num{5094}$ & $\num{196588}$ & $\num{4.71}$ & $\num{13377}$ \\
    \texttt{tw-2010} BFS & $\num{11400}$ & $\num{0.27}$ & $\num{6156}$ & $\num{114356}$ & $\num{2.74}$ & $\num{5154}$ & $\num{192121}$ & $\num{4.61}$ & $\num{13587}$ \\
    \texttt{uk-2007-02} DFS & $\num{13154}$ & $\num{0.11}$ & $\num{16286}$ & $\num{154338}$ & $\num{1.40}$ & $\num{2883}$ & $\num{177945}$ & $\num{1.61}$ & $\num{2248}$ \\
    \texttt{uk-2007-02} BFS & $\num{13155}$ & $\num{0.11}$ & $\num{16287}$ & $\num{156467}$ & $\num{1.42}$ & $\num{2936}$ & $\num{179206}$ & $\num{1.62}$ & $\num{2781}$ \\
    \bottomrule
  \end{tabular}
  }
  \caption{Running time (in milliseconds) and memory usage (in MB) for running breadth-first and depth-first search on both the uncompressed and the compressed representations (both with Zuckerli and WebGraph) of various graphs. We also report the average time (in $\mu s$) to access each adjacency list.\label{tab:perf}}
\end{table*}
\begin{table}[t]
  \centering
  \begin{tabular}{lrrrrrr}
    \toprule
         & \multicolumn{2}{c}{Zuckerli} & \multicolumn{2}{c}{WebGraph} \\
    name & time & memory & time & memory \\
    \midrule
    \texttt{cnr-2000-hc} & $\num{0.03}$ & $\num{5}$ & $\num{0.36}$ & $\num{107}$ \\
    \texttt{cnr-2000} & $\num{0.03}$ & $\num{4}$ & $\num{0.36}$ & $\num{109}$ \\
    \texttt{in-2004-hc} & $\num{0.14}$ & $\num{7}$ & $\num{0.66}$ & $\num{179}$ \\
    \texttt{in-2004} & $\num{0.14}$ & $\num{6}$ & $\num{0.63}$ & $\num{182}$\\
    \texttt{eu-2005-hc} & $\num{0.18}$ & $\num{10}$ & $\num{0.67}$ & $\num{179}$\\
    \texttt{eu-2005} & $\num{0.18}$ & $\num{9}$ & $\num{0.71}$ & $\num{176}$ \\
    \texttt{uk-2002-hc} & $\num{2.20}$ & $\num{54}$ & $\num{4.93}$ & $\num{763}$\\
    \texttt{uk-2002} & $\num{2.16}$ & $\num{65}$ & $\num{5.08}$ & $\num{829}$\\
    \texttt{hw-2009-hc} & $\num{1.35}$ & $\num{64}$ & $\num{2.36}$ & $\num{175}$\\
    \texttt{hw-2009} & $\num{1.33}$ & $\num{65}$ & $\num{2.38}$ & $\num{171}$ \\
    \texttt{tw-2010-hc} & $\num{28.19}$ & $\num{2415}$ & $\num{36.53}$ & $\num{1308}$ \\
    \texttt{tw-2010} & $\num{24.50}$ & $\num{2439}$ & $\num{35.65}$ & $\num{1719}$ \\
    \texttt{uk-2007-02-hc} & $\num{21.32}$ & $\num{445}$ & $\num{46.61}$ & $\num{1701}$\\
    \texttt{uk-2007-02} & $\num{20.84}$ & $\num{573}$ & $\num{49.25}$ & $\num{1617}$ \\
    \bottomrule
  \end{tabular}
  \caption{Running time (in seconds) and memory usage (in MB) for decompressing the graphs sequentially with Zuckerli and with Webgraph.\label{tab:dec}}
\end{table}

We evaluate the performance characteristics of Zuckerli by comparing its running time and memory usage for running depth-first and breadth-first traversals with WebGraph (only for the variants that allow access to single lists), as well as with uncompressed graphs, as a baseline. The running time and the memory usage are reported in Table~\ref{tab:perf}. We also compare the time and memory usage for running a full sequential decompression of the graphs, with results reported in Table~\ref{tab:dec}.

From these comparisons, it emerges that the memory usage for decompression and random access required by WebGraph and Zuckerli is very different, with both methods using less memory in some situations. This can be explained by the different language of the implementation (C++ and Java), as well as the fact that WebGraph uses \emph{lazy iteration} on adjacency lists, to avoid decompressing them fully to memory. While this can in principle be supported by Zuckerli, it was not implemented in this version of the code.

Regarding running time, Zuckerli is often faster than WebGraph. This is due to the fact that Zuckerli requires less memory bandwidth than WebGraph (as it uses less bits for compression), and that it is written in highly optimized C++ code.

Finally, to evaluate the scalability of Zuckerli on multiple cores, we wrote a simple program that computes the sum of all endpoints of all edges of a graph, and we ran it on \texttt{uk-2007-02} using $1$, $2$, $4$, $8$, $16$, $32$ and $64$ cores. The results are shown in Figure~\ref{fig:parallelism}. They show the good scalability of Zuckerli; the speedup is likely limited by memory bandwidth.

%\begin{itemize}
%%%%%%%%%%%%%%%% Added to text
%%\item Large corpus of graphs
%%\item Time for compression/decompression + memory for decompression
%%\item Time and space for running algorithms on random access version (vs uncompressed)
%%\item Effect of window size
%%\item Effect of changing hybrid encoding params
%%\item Effect of disabling optimized heuristic for chain length limit
%%%%%%%%%%%%%%%% Consider adding if time and space allow
%\item (Comparison with ANS for inverted indexes?)
%\item (Effect of changing chunk size?)
%\item (Effect of disabling context modeling?)
%\item (Effect of permuting node order?)
%\end{itemize}

\section{Conclusions}
\label{sec:conclusions}

In this paper, we described Zuckerli, a novel compression algorithm and compressed data structure designed for very large graphs.
By exploiting recent entropy coding techniques, context modeling and improved encoder heuristics based on approximation
algorithms, Zuckerli can achieve significant space savings for compressing web graphs and social networks over
state-of-the-art systems, such as the WebGraph framework. By conducting experiments on a large corpus of web
graphs and social networks, we quantified these savings as roughly $25\%$ on web graphs and roughly $12\%$ on
social networks, both for the full and list decompression use cases. In data compression, this is considered a significant improvement. For example, \texttt{bzip2} is preferred to \texttt{gzip} for file compression when space saving is crucial, because it has $10\sim30\%$ better compression ratios~\cite{burrows1994block}; on the other hand, \texttt{bzip2} is slower and has a larger memory footprint than \texttt{gzip}. Zuckerli achieves similar improvements, but is also faster than Webgraph, with a smaller memory footprint in many cases.
Decompression with Zuckerli is fast, resource-efficient, and scalable. 

% We wish to thank Sebastiano Vigna for useful conversations on the topic. [no! double blind] Paolo Boldi too AC: you're right!! 

\bibliographystyle{abbrv}
\bibliography{main}

%% NOTA : l'appendix e' conteggiato nel numero di pagine. Il PDF finale deve stare nelle 10 pagine.
\appendix

\section{A note on WebGraph's efficiency}\label{sec:apx} %LogGraph ne riporta diversi, siamo interessati a far vedere che webgraph li batte tutti

For completeness, and to motivate our choice of WebGraph as baseline, we refer the comparison already performed by LogGraph~\cite{besta2018log}.
In particular, we report Table~11 from which compares several well known compression techniques including WebGraph. The table shows WebGraph to be consistently more effective than the other techniques, and in the 3 cases where it does not achieve the best compression ratio, it is still very competitive with the best performing method.

\begin{figure}[ht]
    \centering
    \includegraphics[width=\columnwidth]{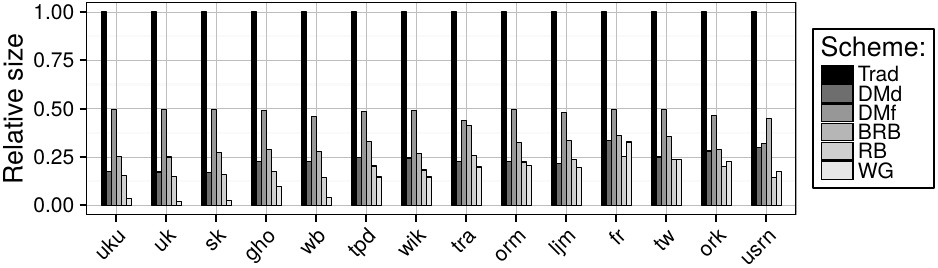}
    \caption{Table~11 from~\cite{besta2018log}, reporting the efficiency of several compression techniques, namely:
    \textit{Trad} : traditional adjacency array,
    \textit{DMd} : Degree-Minimizing \textit{differences}~\cite{besta2018log},
    \textit{DMf} : Degree-Minimizing \textit{full}~\cite{adler2001towards},
    \textit{RB} : Recursive Bisectioning~\cite{blandford2003compact},
    \textit{BRB} : Binary Recursive Bisectioning~\cite{besta2018log}
    \textit{WG} : WebGraph~\cite{boldi2004webgraph}.
    The value reported is the size of the compressed graph relative to \textit{Trad}.
    }
    \label{fig:log-graph-table}
\end{figure}

% \balance

\end{document}